\documentclass[aps,prb,amsmath,amssymb,reprint,superscriptaddress]{revtex4-1}

\usepackage{graphicx}
\usepackage{epstopdf}

\begin{document}


\title{A Quantum Fidelity Study of the Anisotropic Next-Nearest-Neighbour Triangular Lattice Heisenberg Model}


\author{Mischa Thesberg}
\author{Erik S. S{\o}rensen}
\email[]{thesbeme@mcmaster.ca}
 \affiliation{Department of Physics \& Astronomy, McMaster University\\1280 Main St. W., Hamilton ON L8S 4M1, Canada.}


\date{\today}

\begin{abstract}
Ground- and excited-state quantum fidelities in combination with generalized quantum fidelity susceptibilites, obtained from exact
diagonalizations, are used to 
explore the phase diagram of the anisotropic next-nearest-neighbour triangular Heisenberg model.
Specifically, the $J'-J_2$ plane of this model, which connects the $J_1-J_2$
chain and the anisotropic triangular lattice Heisenberg model, is explored using
these quantities.  Through the use of a quantum fidelity associated with the
first excited-state, in addition to the conventional
ground-state fidelity, the BKT-type transition and Majumdar-Ghosh point of the
$J_1-J_2$ chain ($J'=0$) are found to extend into the $J'-J_2$ plane and 
connect with points on the $J_2=0$ axis thereby forming bounded regions in the phase diagram.  
These bounded regions are then explored through the generalized quantum fidelity susceptibilities $\chi_{\rho}$,
$\chi_{120^{\circ}}$, $\chi_D$ and $\chi_{CAF}$ which are associated with the
spin stiffness, $120^{\circ}$ spiral order parameter, dimer order parameter and
collinear antiferromagnetic order parameter respectively.  These quantities are
believed to be extremely sensitivity to the underlying phase
and are thus well suited for finite-size studies.  Analysis of the fidelity
susceptibilities suggests that the $J',J_2 \ll J$ phase of the anisotropic
triangular model is either a collinear antiferromagnet or possibly a gapless disordered phase that is directly connected to
the Luttinger phase of the $J_1-J_2$ chain.  Furthermore, the outer region is
dominated by incommensurate spiral physics as well as dimer order.
\end{abstract}


\maketitle

\section{Introduction}
\label{sec:introduction}
\begin{figure}[t]
\includegraphics[scale=0.55]{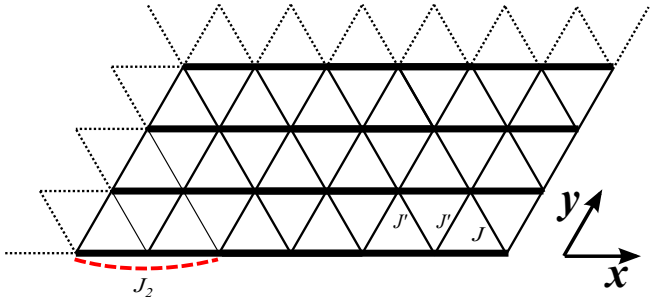}
\caption{\label{fig:Triangular_Lattice_Diagram}  
The anisotropic triangular lattice with next-nearest neighbour interactions.
In this paper $J'$ and $J_2$ are assumed to be ratios of $J$ (i.e. $J=1$).
In the limit $J' \ll 1$ the system can be viewed as a set of weakly coupled
chains.  The next-nearest neighbour
interactions $J_2$ are in the intra-chain direction (red dashed line). A system size is denoted
as $N=W \times L$ corresponding to a system of $W$ chains of length
$L$.  The system size studied here is $4 \times 6$. }

\end{figure}

The study of quantum phase transitions (QPTs), especially those which occur in
two- and one-dimensional systems, remains one of the most active areas of
research in condensed matter physics.\cite{Sachdev_QPT_Book}  Of particular
interest are systems with competition between interactions that cannot be
mutually satisfied.  This behaviour, often arising from  \emph{frustration}, acts
to erode the tendency towards classical orderings and promotes exotic phases
dominated by quantum fluctuations.  Unfortunately, these quantum fluctuations
manifest as highly oscillatory, fermionic field theories.  Such theories cause
Quantum Monte-Carlo (QMC) methods, numerical methods which allow the study of
some of the largest system sizes that are accessible computationally, to fail.
In contrast, Exact Diagonalization (ED) methods, that we employ here, are not affected by the presence
of frustration and can quite generally be applied to lattice models with a finite Hilbert space. 
They are, however, restricted to very small system sizes. 
The use of complimentary methods such as the Density Matrix Renormalization Group (DMRG)
and related methods are therefore also extremely valuable and DMRG results for two-dimensional triangular lattice
models have already been obtained~\cite{Weichselbaum_and_White_DMRG}. However, our focus here is on the
information that can be extracted from ED results in combination with new insights arising from the field of
quantum information.

The numerical identification of QPTs and the classification of their adjoining
quantum phases often involves some \emph{a priori} knowledge about the ordering
of the system and the evaluation of quantities, such as the spin stiffness or
order parameter, which may have poor behaviour or slow/subtle divergences in
small finite systems.  A relatively new quantity, with its origin in the field
of quantum information, has shown promise as a useful numerical parameter for
characterizing QPTs; the quantum fidelity and quantum fidelity
susceptibility.\cite{Quan_Original,Zanardi_Original,Zhou_Original,Connection_FS_and_Dynamic_Structure_Factor}
These quantities have already been successfully employed towards the
identification of QPTs in a number of
systems,\cite{Zanardi_Free_Fermion_Systems,Cozzini_Graphs,Fidelity_and_Matrix_Product_States,Buonsante_Bose_Hubbard,Fidelity_Hubbard,Monte_Carlo_Approach_to_Fidelity,Fidelity_Scaling_1,Fidelity_Susceptibility_XY_Chain,FS_w_TBC_for_1D_Hubbard,
  TFIM_FS_Exact_Solution,Transverse_Longitudinal_FS_2D_TFIM,FS_2D_Square,FS_Anisotropic_Critical_Point,FS_XXZ_Chain_w_DM,FS_2D_XXZ_and_TFIM,FS_Asymmetric_Hubbard_Chain,FS_1D_Transverse_Field_Compass_Model,
  FS_Anisotropic_2D_XYX,Fidelity_and_Spin_Liquids,FS_1D_XXZ_BKT_1,FS_Disordered_XY_Chain,FS_XY_Spin_Chain_w_DM,FS_2D_TFIM_Other,Fidelity_and_Topological_Phases,FS_in_1D_XXZ_BKT_2}
and an excellent review of this approach can be found in
Ref.~\onlinecite{Gu_Review}.  In this paper we will be concerned with
attempts to slightly generalize the notion of the standard fidelity in order
to construct new quantities that can aid in identifying phase transitions in
small systems.  Extensions of the basic fidelity concept are not new, with
prior developments such as the operator fidelity
susceptibility\cite{Operator_Fidelity_Susceptibility} and the reduced
fidelity\cite{Reduced_Fidelity_1,Reduced_Fidelity_2,Reduced_Fidelity_in_Topological_Phases}
having proved fruitful.  
Here we consider two additional variants that have been proposed:
excited-state fidelities~\cite{Second_Eigenvector_Fidelity_J1_J2_Chain} and generalized fidelity susceptibilities~\cite{Thesberg_Generalized_FS,Spin_Current_Fidelity_1D}

The typical quantum fidelity assumes that the Hamiltonian of a system with a QPT can be written in the form
\begin{equation}
H(\lambda) = H_0 + \lambda H_{\lambda},
\end{equation} 
where the phase transition occurs at some critical value of the \emph{driving
parameter} $\lambda$ ($\lambda_c$).  From this perspective the second term is
then seen as the \emph{driving term} and it is entirely responsible for the
phase transition.  The quantum fidelity is then defined as the overlap or
inner-product of the ground-state of a system with another ground-state
determined by a Hamiltonian that is slightly perturbed in the driving parameter
relative to the first:
\begin{equation}
F_0(\lambda,\delta \lambda) = \left\langle \Psi_0 (\lambda) \right. \left\vert \Psi_0 (\lambda + \delta \lambda) \right\rangle,
\end{equation}
where $\Psi_0(\lambda)$ is the ground-state of the Hamiltonian $H(\lambda)$.
In a study by Chen \emph{et al.}\cite{Second_Eigenvector_Fidelity_J1_J2_Chain} of the $J_1-J_2$ chain, a
system we also consider here, it was shown that a fidelity based not on the ground-state
but the first excited-state,
\begin{equation}
F_1(\lambda, \delta \lambda) = \left\langle \Psi_1 (\lambda) \right. \left\vert \Psi_1 (\lambda + \delta \lambda) \right\rangle,
\end{equation}
could be a potentially valuable quantity.  Here we call such a fidelity an excited-state fidelity.

From the quantum fidelity one can calculate the quantum fidelity susceptibility, defined as
\begin{equation}
\chi_{\lambda} = \frac{2(1 - F_0(\lambda)}{\delta \lambda^2}.
\end{equation}
However, in a previous work~\cite{Thesberg_Generalized_FS} it was shown that this definition could
be extended by considering other types of perturbations beyond a perturbation
in the driving parameter.  Specifically, it is often useful to construct
generalized fidelity susceptibilities associated with the order parameters of
common orderings.\cite{Spin_Current_Fidelity_1D} 

Our goal here is to explore the phase-diagram of the anisotropic next-nearest-neighbor
triangular lattice model (ANNTLHM). This model connects the $J_1-J_2$ chain ($J'=0$) 
with the anisotropic triangular lattice Heisenberg model (ATLHM) ($J_2=0$). 
The phase diagram of the ATLHM for $J'\ll 1 $ and accordingly of the ANNTLHM for $J',J_2\ll 1$
has proven exceedingly difficult to determine and it appears that several possible phases {\it very}
closely compete.

The $J_1-J_2$ chain has the Hamiltonian 
\begin{equation}
\label{eq:J1_J2_Hamiltonian}
H_{J_1-J_2} = \sum_{\mathbf{x}} \hat{S}_{\mathbf{x}} \cdot \hat{S}_{\mathbf{x+1}} + J_2 \sum_{\mathbf{x}} \hat{S}_{\mathbf{x}} \cdot \hat{S}_{\mathbf{x+2}}
\end{equation}
where $J_2$ is understood to be the ratio ($J_2=J_2'/J_1'$) of the 
next-nearest neighbour ($J_2'$) and nearest-neighbour ($J_1'$) interaction
constants. It is a system which has been well studied; both through field
theoretic approaches,\cite{Haldane_J1_J2,Affleck_Field_Theory_J1_J2_Impurity}
and through numerical approaches like exact
diagonalization,\cite{Eggert,J1_J2_ED_1} and DMRG.\cite{DMRG_2,DMRG_3}  These
studies have revealed the existence of a rich phase diagram for the $J_2>0$
region.  For $J_2<J_2^c \sim 0.241$\cite{Eggert} the
system exhibits a disordered Luttinger liquid phase characterized by
quasi-long-range order (i.e. algebraic decay of spin-spin correlations) and no
excitation gap.  At $J_2^c$ an energy gap opens and for
$J_2^c<J_2$ dimerization sets in and correlations become
short-ranged.  
At the so called Majumdar-Ghosh (MG) point $J_2^{MG}=J/2$ the
ground-state of the system is known exactly and with periodic boundary conditions it is
exactly two-fold degenerate even for finite systems, a fact that is important for our study.
Slightly away from the MG point the degeneracy is lifted for finite systems with an exponentially small
separation between the odd and even combinations of the two possible dimerization patterns.
The correlation length of the system reaches a minimum at the MG point.\cite{Andreas_Paper}  
The MG point can also be identified as a {\it disorder} point marking the onset of incommensurate correlations in real-space occuring for $J_2>J_{MG}$.
The incommensurate effects occuring for $J_2>J_2^{MG}$ are short-ranged and the system remains dimerized for any finite $J_2>J_2^c$.
Of particular importance to us here is the Luttinger liquid-dimer
transition at $J_2^c$, which is known to be in the BKT universality class and
difficult to detect numerically, and the onset of incommensurate correlations
at the MG point $J_2^{MG}$. As we shall show here it is possible to track these points into the $J'-J_2$ plane
of the ANNTLHM.

The ATLHM (see Fig.~\ref{fig:Triangular_Lattice_Diagram}) is described by the Hamiltonian 
\begin{align}
\label{eq:ATLHM_Hamiltonian}
H_{\Delta}= \sum_{\mathbf{x},\mathbf{y}} \hat{S}_{\mathbf{x},\mathbf{y}}\hat{S}_{\mathbf{x}-1,\mathbf{y}} + J' \sum_{\mathbf{x},\mathbf{y}} \hat{S}_{\mathbf{x},\mathbf{y}} \cdot \left( \hat{S}_{\mathbf{x},\mathbf{y+1}} + \hat{S}_{\mathbf{x}-1,\mathbf{y+1}} \right),
\end{align}
where, like $H_{J_1-J_2}$, the coupling constant $J'$ is taken to be the ratio
of the two exchange constants corresponding to the two different exchange
terms.  The phase diagram of this system for $J'<1$ has proven extremely hard to 
determine and many aspects are still undecided.  Early
interest in this system was fuelled by initial theoretical and numerical
studies\cite{Weng_Weng_Bursill_ED,Becca_Sorella_ED,Chung_Marston_LSW_HATM}
which suggested the existence of a  two-dimensional spin liquid phase
for $J' \ll 1$.  This was especially exciting since the ATLHM is believed to be
an accurate description of a number of real experimental materials, such as:
the organic salts $\kappa-$(BEDT-TTF)$_2$Cu$_2$(CN)$_3$
\cite{NMR_Organic_Salt,Sachdev_Organic_Salts,Organic_Salts_Hamiltonian_Work}
and
$\kappa-$(BEDT-TTF)$_2$Cu$_2$[N(CN)$_2$];\cite{Organic_Salts_Hamiltonian_Work}
and the inorganic salts
Cs$_2$CuCl$_4$,\cite{Coldea_Tsvelik_Neutron_Scatter_CsCuCl,Balents_Initial_CsCuCl,Coldea_Initial,Coldea_Hamiltonian_Work,Coldea_Zheng}
and Cs$_2$CuBr$_4$.\cite{Coldea_Zheng,Tanaka_CsCuBr} However, later theoretical
studies would suggest that experimental results on Cs$_2$CuCl$_4$ could be
explained within the paradigm of a less exotic quasi-one-dimensional spin
liquid.\cite{Kohno_Starykh_Balents_Nature,Balents_Nature}  This too gave way to
a number of recent renormalization group studies which suggest that the $J' \ll
1$ region is not a spin liquid at all but rather that next-nearest chain
antiferromagnetic interactions and order-by-disorder give rise to a collinear
antiferromagnetic (CAF) ordering.\cite{Balents_RG,Kallin_Ghamari_RG}  In prior work~\cite{Thesberg_HATM_Twist},
we have studied this system through the use of twisted boundary conditions which alleviate
some of the finite-size issues associated with incommensurate correlations. The application of twisted
boundary conditions suggests
the existence of incommensurate spiral ordering for $J' \sim 1$ giving
way, after a phase transition, 
to a new phase dominated by antiferromagnetism albeit with
short-range incommensurate spiral correlations.  In Ref.~\onlinecite{Thesberg_HATM_Twist}
a rough thermodynamic limit extrapolation suggested the new phase was
gapless, though whether a true collinear antiferromagnetic ordering, as
suggested by Balents \emph{et al.},\cite{Balents_RG} emerged could not be
definitively determined.

Here we are concerned with the application of excited-state fidelity and generalized fidelity susceptibility techniques to the more
general Hamiltonian (ANNTLHM) including a next-nearest neighbor coupling along the chains:
\begin{eqnarray}
\label{eq:The_Hamiltonian}
H (J',J_2) &=& \sum_{\mathbf{x},\mathbf{y}} \hat{S}_{\mathbf{x},\mathbf{y}}\hat{S}_{\mathbf{x-1},\mathbf{y}} \nonumber \\
 &+& J' \sum_{\mathbf{x},\mathbf{y}} \hat{S}_{\mathbf{x},\mathbf{y}} \cdot \left( \hat{S}_{\mathbf{x},\mathbf{y+1}} + \hat{S}_{\mathbf{x-1},\mathbf{y+1}} \right) \nonumber \\
&+&  J_2 \sum_{\mathbf{x}} \hat{S}_{\mathbf{x}} \cdot \hat{S}_{\mathbf{x-2}}.
\end{eqnarray}
As $J', J_2$ are varied the ANNTLHM interpolates between the $J_1-J_2$ chain ($J'=0$ and the ATLHM ($J_2=0$) through the
creation of a $J'-J_2$ plane (see Fig. ~\ref{fig:Triangular_Lattice_Diagram}).
To our knowledge such a general system has only been studied field
theoretically\cite{Balents_RG,Kallin_Ghamari_RG}  and is believed to exhibit
the CAF order discussed previously for small $J'$ and $J_2$ before transiting
to spiral ordering for large $J'$, small $J_2$, and dimer ordering for large
$J_2$, small $J'$.  We will now more thoroughly introduce and define the
excited-state fidelity and generalized fidelity susceptibilities.

\section{Excited-State Fidelities}
\begin{figure}[t]
\includegraphics[scale=0.5]{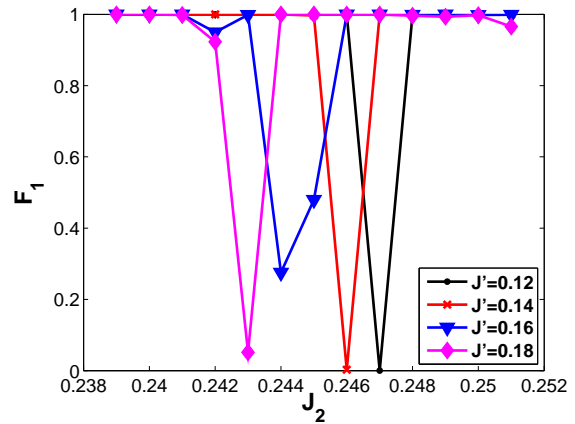}
\caption{\label{fig:Sample_F_1_Sweep} 
(Colour available on-line) The first excited-state fidelity vs. $J_2$ for
values of $J'$ of 0.12 (black circles), 0.14 (red crosses), 0.16 (blue
triangles) and 0.18 (magenta diamonds).  The sharp downward spikes represent
a level crossing in the excited state which was shown by Chen  \emph{et al.} to
signify the BKT-type transition point of the $J_1-J_2$
chain.\cite{Second_Eigenvector_Fidelity_J1_J2_Chain}}
\end{figure}
In the context of the quantum fidelity it is sometimes useful to consider a quantum
phase transition as a result of a level crossing in the ground- or
excited-states as a function of the driving parameter
$\lambda$.\cite{Gu_Review} 
This is a perspective that has proven useful
for the study of a class of one-dimensional
models\cite{Excited_State_Level_Crossing} and can be partly motivated by the
consideration that quantum phase transitions are the result of sudden
reconfigurations of the low-lying energy spectrum of a system.

Motivated by this viewpoint it was shown in
Ref.~\onlinecite{Second_Eigenvector_Fidelity_J1_J2_Chain} (see also Ref.~\onlinecite{Eggert}) that the BKT-type
transition in the $J_1-J_2$ can be detected, in finite-systems, by locating a level
crossing in the \emph{first} excited-states.  Thus, the
determination of the transition point at $J_2 \sim 0.24$ was possible by
constructing a fidelity, $F_1$, not of the ground-state but of the first excited-state.
Using this excited-state fidelity it was
demonstrated~\cite{Second_Eigenvector_Fidelity_J1_J2_Chain} that an abrupt drop
in $F_1$ as a result of the excited state level crossing occurs at the BKT transition point.  
Here, we use the same
fidelity to follow the behaviour of this transition as it extends into
the $J'-J_2$ plane.
We note that, from a numerical perspective, it is considerably more convenient to monitor $F_1$ rather
than the associated level crossing since the latter would require an intricate analysis of several of the low-lying states.

A careful analysis of
Ref.~\onlinecite{Second_Eigenvector_Fidelity_J1_J2_Chain}, specifically Fig. 5
there-in, also indicates the presence of a \emph{ground-state} level crossing
at the Majumdar-Ghosh point\cite{Majumdar_Ghosh_Paper} for finite-systems as mentioned above.
This crossing, which occurs where it is known no actual phase transition occurs in the thermodynamic limit, could be
detected by the ground-state fidelity ($F_0$) and coincides with the
onset of short-range incommensurate correlations in real space even though no long-range spiral order develops.
For a two-dimensional system such as the ATLHM it is known that spiral order occurs close to $J'=1$ and it is
is then also of considerable interest to see if it is possible to track this level crossing through the $J'-J_2$ plane and
what bearing, if any, it has on the physics of the ANNTLHM.

To this end, the ground-state and first excited-state of the ANNTLHM were
calculated for a $4 \times 6$ triangular lattice with periodic boundary
conditions using a parallel, Lanczos, exact diagonalization code as outlined by
Lin \emph{et al.}\cite{Lin_ED_Paper}  Total-$S^z$ symmetry was invoked and
numerical errors in ground-state eigenenergies are estimated to be on the order
of $10^{-10}$.  Numerical errors in the first excited-state energies, as is a
drawback of the Lanczos method, are considered to be higher by an order of
magnitude.  It is worth noting that when constructing the excited-state
fidelity, and thus solving for the eigenvector of the first excited-state, the
difficulty in the Lanczos method of ghost eigenvalue formation is exacerbated
and special care must be taken to throw out erroneous results.

\begin{figure}[t]
\includegraphics[scale=0.5]{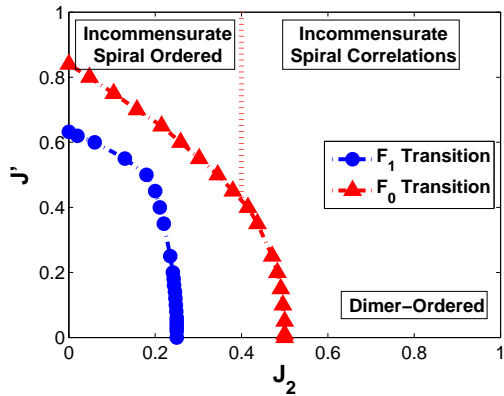}
\caption{\label{fig:Fidelity_Phase_Diagram} 
(Colour available on-line) The  phase diagram of the $J'-J_2$ plane of the ANNTLHM with
regards to the BKT-type transition of the $J_1-J_2$ chain, identified by the
excited-state fidelity $F_1$, in blue (circles) and the crossover associated
with the onset of short-range incommensurate spiral correlations, identified
  by $F_0$, in red (triangles).  All results are for a $4\times 6$ system. 
  The dotted red line represents a rough
  estimate, evidenced by the dimer fidelity susceptibility, of the region where
  long-range spiral order transitions to short-range incommensurate spiral correlations transition with possible dimer order.
  It is derived from the data in Fig.~\ref{fig:Chi_D_Jp_Sweep}. }

\end{figure}

Once the ground-state and first excited-state eigenvectors were obtained
numerically, $F_0$ and $F_1$ were constructed as a function of $J_2$.  A typical
tracking of the drop in $F_1$ is shown for various values of $J'$ between $0.12$
and $0.18$ versus $J_2$ in Fig.~\ref{fig:Sample_F_1_Sweep}.  The path of the transition in
$F_0$ is traced in a similar manner.  As mentioned above, we calculate $F_0, F_1$ and therefore only
gain indirect information about an assoicate level crossing. However, a further examination of the energy
spectrum characteristics which produce the spike in $F_0$ reveals that it is
either due to a ground-state level crossing which persists into the $J'-J_2$
plane or an extremely close avoided level crossing.   The resulting phase
diagram implied by this finite system is shown in
Fig.~\ref{fig:Fidelity_Phase_Diagram}. All results are obtained using a $4\times 6$ system.

One can see that both transitions, when followed, persist well into the
$J'-J_2$ plane and ultimately terminate along the $J_2=0$ line.  This line
corresponds to the ATLHM and it is therefore fruitful to consider their
interpretation within the context of that system.  However, a thorough
consideration with respect to the nearest-neighbour triangular model will be
left to section \ref{sec:Results_and_Discussion}, after the introduction of the
generalized fidelity susceptibilities.  For now it is sufficient to realize
that the level-crossing observed at the Majumdar-Ghosh point in the $J_1-J_2$ chain
ultimately connects with the parity transition observed in previous numerical
investigations of the ATLHM.\cite{Weng_Weng_Bursill_ED,Becca_Sorella_ED}  In
Ref.~\onlinecite{Thesberg_HATM_Twist} we studied the same system 
through the use of twisted boundary conditions, which allow a more
natural treatment of incommensurate behaviour, and in it was found
that, although a transition does occur, this parity transition is an
unphysical artefact of a finite-sized system with periodic boundary conditions.
The same conclusion was arrived at in the DMRG study of Weichselbaum
and White.\cite{Weichselbaum_and_White_DMRG}  Thus, it seems that both in the
$J_1-J_2$ chain (where it is known that incommensurate correlations arise past
the disorder (MG) point) and in the ANNTLHM this transition may indicate the onset of incommensurate physics.

Using ground-state and excited-state fidelities we have thus demarcated a phase
diagram in the $J'-J_2$ plane shown in Fig.~\ref{fig:Fidelity_Phase_Diagram}.  
It is clear tht the quantities $F_0$ and $F_1$ are useful tools
for determining the phase diagram.  However, equally important as the
\emph{location} of QPTs is the nature of the adjacent quantum phases.
It is possible to extend the fidelity approach, through the
introduction of generalized fidelity susceptibilities, to aid in the 
identification of the phase in each region that has been found so far. These
susceptibilities will now be introduced.

\section{Generalized Quantum Fidelity Susceptibilities}

In the previous section we showed the simplicity with which quantum phase
transitions driven by level crossings, either in the ground-state or low-lying
excited-states, can be identified and traced with the quantum fidelity (when
    generalized to the overlap of excited-states).   Once the location of QPTs
within phase space have been charted often the next task, when encountering a
system of interest, is the identification of the various phase regions.
Ideally one would like to be able to associate an order parameter, local or
not, with each demarcated phase (or none for a disordered phase). 

It has been shown by Zanardi \emph{et al.}\cite{Information_Theoretic_Approach}
and Chen \emph{et al.}\cite{Connection_FS_and_Ground_State_Derivatives}, that
there is a close connection between a fidelity susceptibility and the second
derivative of the ground-state energy with respect to the ``driving parameter''
with which  the fidelity susceptibility is constructed:
\begin{eqnarray}
\chi &=& \sum_n \frac{\left\vert \left\langle \Psi_n \right\vert H_I \left\vert \Psi_0 \right\rangle \right\vert^2}{\left( E_0(\lambda) - E_n(\lambda) \right)^2}, \nonumber \\
\frac{\partial^2 E_0(\lambda)}{\partial \lambda^2} &=& \sum_n \frac{2 \left\vert \left\langle \Psi_n \right\vert H_I \left\vert \Psi_0 \right\rangle \right\vert^2}{\left( E_0(\lambda) - E_n(\lambda) \right)}. \nonumber
\end{eqnarray}
As can be seen, the fidelity susceptibility has a higher power in the denominator and
is therefore expected to have a higher sensitivity.  It is important to note
that this relationship holds true even if the ``driving'' parameter and
Hamiltonian ($\lambda$ and $H_{\lambda}$) are not actually the terms that drive
the phase transition.  In
Ref.~\onlinecite{Thesberg_Generalized_FS}, it was demonstrated that for the
$J_1-J_2$ chain the different phases can be identified through the use of an appropriately
constructed generalized fidelity susceptibility.

When adopting this approach one begins by identifying all the potential phases
that one suspects might exist within the phase diagram under study.  The
primary task is then to construct a fidelity susceptibility for each of these
phases which has a similar connection to the order parameter susceptibility of
that phase that the regular (i.e. $\lambda$ \emph{is} the driving parameter)
  fidelity susceptibility has with the ground-state derivatives.  It is then expected
 that such a generalized fidelity susceptibility will exhibit the same behaviour as the
  order parameter susceptibility, going to infinity when in the associated
  phase and zero when outside it in the thermodynamic limit, but with increased
  sensitivity in finite systems.

As has been discussed, the $J_1-J_2$ chain studied in
Ref.~\onlinecite{Thesberg_Generalized_FS} serves as a limiting case of the
ANNTLHM as $J' \rightarrow 0$.  Thus, all the fidelity susceptibilities
constructed in Ref.~\onlinecite{Thesberg_Generalized_FS} find use here, once generalized to two dimensions.
To these susceptibilities ($\chi_{\rho}$, $\chi_D$, $\chi_{CAF}$) have been
added the new susceptibility $\chi_{120^{\circ}}$ which is designed to capture
the incommensurate spiral phase of the $J'\sim 1$ region.  We will now
explicitly describe the construction of each of these susceptibilities.

\begin{figure}[t]
\includegraphics[scale=0.5]{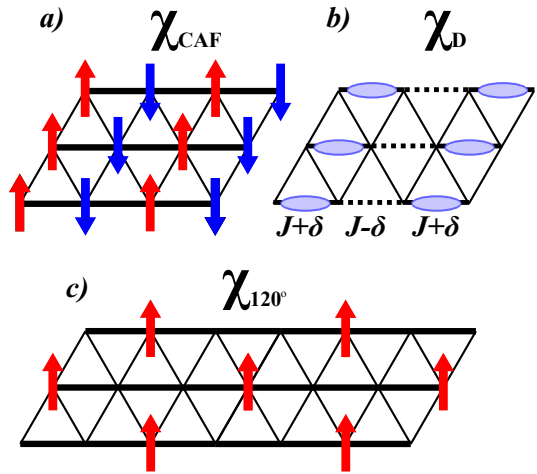}
\caption{\label{fig:Chi_Diagrams}  Diagrams of the perturbing term which define
the generalized fidelity susceptibilities $\chi_{CAF}$, $\chi_D$ and
$\chi_{120^{\circ}}$, respectively.  $\chi_{CAF}$, shown in a), is defined by a
fidelity whose perturbed Hamiltonian is one with an infinitesimal staggered
magnetic field in the $S^z$ direction added according to the illustrated
pattern. $\chi_D$, shown in b), is defined by a perturbation in the
intra-chain, nearest-neighbour, exchange interaction with alternating bonds
having their exchange constant modified by a $\pm \delta$.
$\chi_{120^{\circ}}$, shown in c), is a rough probe of spiral order close to
that known to exist at $J'=1,J_2=0$ and corresponds to an upward magnetic field
on every third site, corresponding to a spiral phase whose ordering has a
wavelength of three sites.  The omission of in-plane fields on the remaining
sites is to maintain the conservation of total-$S^z$ in the system Hamiltonian
which improves numerics. }
\end{figure}

\subsubsection{The CAF Fidelity Susceptibility, $\chi_{CAF}$}

The collinear antiferromagnetic susceptibility is the natural two-dimensional
extension of the antiferromagnetic fidelity susceptibility ($\chi_{AF}$)
introduced in Ref.~\onlinecite{Thesberg_Generalized_FS}.  It is constructed
by choosing a perturbing Hamiltonian representing a staggered magnetic field
which tiles the lattice (See Fig.~\ref{fig:Chi_Diagrams}a):
\begin{equation}
\lambda H_{CAF}= \lambda \sum_{\mathbf{y}=0}^{W-1} \sum_{\mathbf{x}=0}^{L-1} (-1)^{\mathbf{x}} S^z_{\mathbf{x},\mathbf{y}}.
\end{equation}
The generalized fidelity susceptibility associated with this perturbation is then
\begin{equation}
\chi_{CAF}= \frac{2(1-F(\lambda,J',J_2) )}{\lambda^2}
\end{equation}
where $F(\lambda,J',J_2) $ is given by
\begin{equation}
F(\lambda,J',J_2) = \left\vert \left\langle \Psi_0( 0, J' ,J_2) \right\vert \left. \Psi_0 (\lambda , J' , J_2) \right\rangle \right\vert.
\end{equation}
\begin{figure}[h]
\includegraphics[scale=0.5]{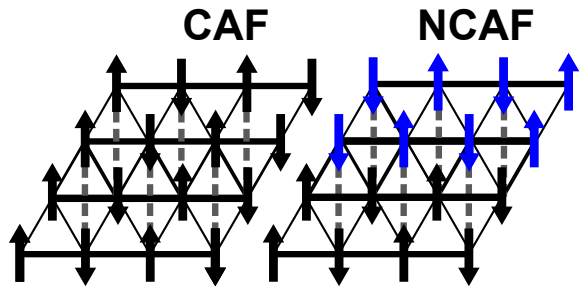}
\caption{\label{fig:CAF_vs_NCAF_Diagram}  A diagram comparing the staggered
  magnetic field arrangement of collinear antiferromagnetic (CAF) vs.
    non-collinear antiferromagnetic (NCAF) orderings.  The key difference is
    whether next-nearest-chain correlations are antiferromagnetic or
    ferromagnetic.  Field theory work suggests that CAF correlations force an
    ordered state for $J' \ll 1$.\cite{Balents_RG}  Generalized fidelity
    susceptibilities were constructed for both CAF and NCAF fields and
    $\chi_{CAF}$ was found to be greater than $\chi_{NCAF}$, though only by a
    tiny, but meaningful, factor of 0.001$\%$.} \end{figure}

As already mentioned, previous
work\cite{Balents_RG,Thesberg_HATM_Twist,Kallin_Ghamari_RG} on the ANNTLHM has
emphasized the important physical difference between antiferromagnetic tilings
where next-nearest chain interactions are antiferromagnetic and ferromagnetic
as indicated by the dashed lines in Fig. ~\ref{fig:CAF_vs_NCAF_Diagram}.  In this work we denote the
ferromagnetic case as NCAF (non-collinear antiferromagnetic) ordering and the
antiferromagnetic as CAF.  Thus, we see that the tiling presented in Fig.~\ref{fig:Chi_Diagrams}a is
indeed $\chi_{CAF}$.  Later we will compare the value of this susceptibility
with that for a susceptibility with NCAF ordering, $\chi_{NCAF}$:
\begin{equation}
\lambda H_{NCAF}=  \lambda \sum_{\mathbf{y}=0}^{W-1} \sum_{\mathbf{x}=0}^{L-1} (-1)^{ \lfloor j/2 \rfloor} S^z_{\mathbf{x},\mathbf{y}}  
\end{equation}
where $\lfloor x \rfloor$ represents the \emph{floor} (i.e. rounded down to the
nearest integer) of $x$.  Thus, the additional term switches the ordering
every \emph{two} chains and thus produces an NCAF tiling as shown in the right panel of Fig. ~\ref{fig:CAF_vs_NCAF_Diagram}.
 
The procedure for the calculation of $\chi_{CAF}$ then simply amounts to
solving for the ground-state of the system when $\lambda=0$ and again when
$\lambda$  is some small number. The  inner product of the two resulting wave-functions then yields the
fidelity.  This fidelity is then converted to a susceptibility.  We contend
that this fidelity susceptibility will have the same properties as the order
parameter susceptibility of a collinear-antiferromagnetic phase but with an
increased sensitivity, making it more useful for the small system sizes
available through ED.  \\

\subsubsection{The Dimer Fidelity Susceptibility, $\chi_D$}
The dimerized susceptibility presented in
Ref.~\onlinecite{Thesberg_Generalized_FS} is easily extended to two-dimensions.
This susceptibility, dictated by the perturbing Hamiltonian
\begin{equation}
\delta H_{D} =  \delta \sum_{\mathbf{y}=0}^{W-1} \sum_{\mathbf{x}=0}^{L-1} (-1)^{\mathbf{x}} S^z_{\mathbf{x},\mathbf{y}} S^z_{\mathbf{x+1},\mathbf{y}},
\end{equation}
corresponds to a dimer tiling \emph{along} chains (here we use $\delta$ rather
than $\lambda$ to emphasize the similarity to the classic dimerization
operator).  One could construct a similar susceptibility which assumes
dimerization in the $J'$ direction.  However, such a tiling was found to be far
less important, this could have been expected \emph{a priori} since the energy
benefit of such inter-chain singlet formation is less than that for intra-chain
singlets.  It is also worth noting that, in principle, one could have two
different tilings \emph{with} intra-chain singlets corresponding to a vertical
(i.e. along $(0,1)$) and diagonal (i.e. along $(1,1)$) stacking.  However, no
numerical difference was found between these two possibilities.

As before, a quantum fidelity susceptibility, $\chi_{D}$ is constructed from the fidelity associated
with this perturbing Hamiltonian and we take it to be related to the order
parameter susceptibility of a dimerized phase.\\

\subsubsection{The Spin Stiffness Fidelity Susceptibility, $\chi_{\rho}$}

The spin stiffness is defined as 
\begin{equation}
\rho(L) = \left.  \frac{\partial^2}{\partial \theta^2}  \frac{E_0(\theta)}{L}\right\vert_{\theta=0}
\end{equation}
where $E_0(\theta)$ is the ground-state energy as a function of a twist $\theta$ applied at every bond:
\begin{eqnarray}
\label{Eq:stiffness_transformation}
H_0 & \rightarrow & H_{\rho} \nonumber \\
 \mathbf{S}_{i} \cdot \mathbf{S}_{j} &\rightarrow & S^{z}_i S^z_j + \frac{1}{2} \left(S^+_i S^-_j e^{i \theta} + S^{-}_i S^{+}_j e^{-i \theta} \right).
\end{eqnarray}
It has proven to be a useful quantity in the exploration of quantum phase
diagrams for it can be taken as a measure of the level of spin order exhibited
by a phase.  In a quasi-long-range ordered system like the Heisenberg chain it
is known to take a non-zero value in the thermodynamic
limit,\cite{Spin_Stiffness_1,Spin_Stiffness_2} the same is true for a system
with spin ordering.  It would be zero in a non-spin ordered system in the
thermodynamic limit.  The behaviour in finite systems can be less
straightforward though it can be said that the sensitivity of a system with
respect to an infinitesimal twist can provide valuable information as to the
strength of spin-correlations and tendency to order, even in small systems.  To
benefit from the information stored in a quantity like the spin stiffness while
maintaining the sensitivity gains afforded by a fidelity susceptibility we then
construct a spin stiffness fidelity susceptibility, $\chi_{\rho}$.  Such a
susceptibility is constructed, not by the usual addition of a perturbing conjugate
field, but through the transformation Eq.~(\ref{Eq:stiffness_transformation}) of the
system Hamiltonian.  One then calculates the overlap of the ground-state of the
Hamiltonian with no twist and with an infinitesimal twist in order to construct
the appropriate fidelity.  Although this does not strictly follow the same form
as the other fidelities one could expand the exponential in $\theta$ to obtain
an $H=H^{(0)} + \theta H_{\theta}^{(1)} + \theta^2 H^{(2)}_{\theta}$ form.  As
is discussed in more detail in Ref.~\onlinecite{Thesberg_Generalized_FS}, one
can then identify $H_{\theta}^{(1)}$ as a spin current operator and
$H_{\theta}^{(2)}$ as a spin kinetic energy term (see also Ref.~\onlinecite{Spin_Current_Fidelity_1D}). However, the numerical
difference between the exponential and Taylor expanded forms was found to be
negligible and thus in this paper we will merely treat things as an
exponential.

We then take the fidelity susceptibility constructed from this spin stiffness
fidelity to be a sensitive measure of spin ordering in a probed phase.

\subsubsection{The 120 Degree Fidelity Susceptibility, $\chi_{120^{\circ}}$}

For the isotropic case of $J'=1$ ($J_2=0$) the triangular lattice is known to
exhibit a spiral phase with a wavevector of $2 \pi /3 $ or
$120^{\circ}$.\cite{Isotropic_Triangular_120_Order}  As $J'$ becomes less than
1 this spiral order persists, although with incommensurate wavevectors.
However, associating a susceptibility with an incommensurate ordering is not
feasible without knowledge of the $q$-vector beforehand.  One could invoke
estimates of these incommensurate ordering vectors obtained in both the
prior studies~\cite{Weichselbaum_and_White_DMRG,Thesberg_HATM_Twist}
and construct a separate
fidelity susceptibility for each value of $J'$.  However, here we employ a
simpler, though likely less accurate, approach by defining a generalized
fidelity susceptibility for the $120^{\circ}$ ordering case only.  In the limit
of $J'\rightarrow 0$ the classical system will be antiferromagnetically ordered
and thus we can expect, in this limit, that $\chi_{CAF}$ can correctly identify
ordering here.  We thus expect a transition from an ordering of wavelength
three to an incommensurate ordering with approximate wavelength of two for
small systems.  Therefore, we can expect a generalized fidelity susceptibility
associated with both these limits (i.e. $\chi_{120^{\circ}}$ and $\chi_{CAF}$)
  to provide valuable information about the ordering across the $J_2=0$ phase
  diagram and outwards.

In order to construct $\chi_{120^{\circ}}$ an $S^z$ magnetic field is placed on
every third site along a chain (see Fig.~\ref{fig:Chi_Diagrams}c) while all
other sites were left unaffected.  The reason that no magnetic field is placed
on the other sites is that the addition of magnetic fields in the $S^x-S^y$
plane would break total-$S^z$ symmetry and significantly complicate numerics.
Thus, $\chi_{120^{\circ}}$ is constructed in an almost identical fashion to
$\chi_{CAF}, \chi_{NCAF}$ except for the location of the perturbing magnetic
fields.

\subsubsection{Comparing Generalized Susceptibilities}

The fidelity susceptibilities constructed here are the result of significantly
different perturbations with different scaling and absolute magnitude i.e.
$\chi_{CAF}$ and $\chi_{NCAF}$ see the addition of 24 perturbing fields for $N=4 \times 6$
where as $\chi_{120^{\circ}}$ sees only the addition of 8.  It is therefore sensible
to compare $\chi_{CAF}$, $\chi_{NCAF}$ with $3\times\chi_{120^\circ}$.
However, there is no obvious way to quantitatively compare these fidelity susceptibilites
to $\chi_D$ and $\chi_\rho$ for a single system size. Instead a detailed finite-size scaling analysis
of the different suceptibilities should be done. For the two-dimensional systems we are considering
here it is not possible to perform such a finite-size scaling analysis using ED techniques.
In fact, when plotting the susceptibilities
arbitrary multiplicative coefficients will be added in front of $\chi_\rho$ ($\times 3$) 
and $\chi_{120^\circ}$ ($\times 30)$ in order to produce graphs with
all susceptibilities visible.  It is therefore only qualitative comparisons
that can be made between these new quantities.  However, as we will see, this
qualitative behaviour tends to be quite drastic and illuminating and thus
provides valuable information about the phase diagram of any system under
consideration.

\section{Results and Discussion}
\label{sec:Results_and_Discussion}

\subsection{The $J_1-J_2$ Chain ($J'=0$)}
\begin{figure}[t]
\includegraphics[scale=0.5]{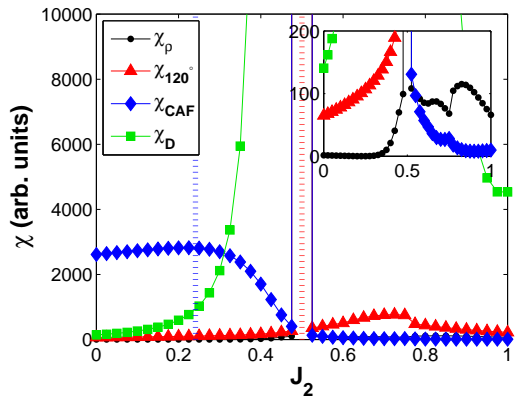}
\caption{\label{fig:FS_Plots_Jp_0p0} (Colour available on-line)  The values of
the generalized fidelity susceptibilities $\chi_{\rho}$ (black circles),
$\chi_{120^{\circ}}$ (red triangles), $\chi_{CAF}$ (blue squares), $\chi_D$
(magenta diamonds) as a function of $J_2$ for $J'=0$ (i.e. the $J_1-J_2$
chain).  All results are for a 24 site $J_1-J_2$ chain.
Also shown are the locations of the transition detected by the
excited-state fidelity $F_1$ (dotted blue line) and the transition detected by
the ground-state fidelity $F_0$ (dotted red line).  
$\chi_{\rho}$ has been scaled by a
factor of three and $\chi_{120^{\circ}}$ has been scaled by a factor of thirty.
All other susceptibilities have not been scaled. The inset shows the same data but with a different y-axis.}
\end{figure}

In order to interpret generalized fidelity susceptibility data in the $J'-J_2$
plane it is prudent to begin in the limit where things are well understood.  In
this system the $J'=0$ case is such a limit for the system then reduces to the
well
studied\cite{J1_J2_ED_1,Bursill_Parkinson_Important_Numerical_J1_J2,DMRG_2,DMRG_3}
$J_1-J_2$ chain.   A plot of $\chi_{\rho}$, $\chi_D$, $\chi_{CAF}$ and
$\chi_{120^{\circ}}$ ($\delta \lambda = 10^{-4}$) is shown in
Fig.~\ref{fig:FS_Plots_Jp_0p0} for a $24$ site $J_1-J_2$ chain as a function of $J_2$.
As such  this data amounts to an extension of the data found in
Ref.~\onlinecite{Thesberg_Generalized_FS}.

For $J_2< 0.2411=J_2^c$ the system is in the
spin-liquid Heisenberg phase marked by quasi-long-range order (i.e. algebraic
    decay of correlation functions to zero with spin separation), a non-zero
spin stiffness,\cite{J1_J2_Spin_Stiffness,Sorensen_XXZ_Chain} and a gapless
excitation spectrum.  Beyond this phase the system is found to develop a gap for
$J_2>J_2^c$.  At the
Majumdar-Ghosh point, $J_2=1/2=J_2^{MG}$, the system, in the thermodynamic
limit, is a perfect superposition of two dimerized states and the ground-state
is known.\cite{Majumdar_Ghosh_Paper} The MG point is a disorder point and for $J_2>J_2^{MG}$
incommensurate effects appear in the real-space correlations.
The ability of generalized fidelities to identify and
characterize the $J_2 < J_2^{MG}$ region and specifically the $J_2=J_{2}^c$
BKT-type transition was established in Ref.~\onlinecite{Thesberg_Generalized_FS}
and thus that analysis will not be repeated here.  

For $J_2<J_2^c$ the dominant fidelity susceptibility is $\chi_{CAF}$, 
associated with the antiferromagnetic correlations in the
Luttinger phase.  (For the $J_1-J_2$ chain $\chi_{CAF}$ used here is identical to $\chi_{AF}$ 
discussed in Ref.~\onlinecite{Thesberg_Generalized_FS}).

For $J_2>J_2^c$, $\chi_{CAF}$ dramatically decreases while $\chi_D$ becomes dominant signalling
the onset of dimer order.
The distinctive behavior of
$\chi_D$ for $J_2>J_2^c$ is reminiscent of the behaviour of the dimer order parameter, whose
numerically calculated value can be found in Fig. 5 in
Ref.~\onlinecite{Bursill_Parkinson_Important_Numerical_J1_J2} and Fig. 8 in
Ref.~\onlinecite{DMRG_3}, albeit with increased sensitivity.  

Looking at Fig.\ref{fig:FS_Plots_Jp_0p0} it is also clear that there is an abrupt
behavior at $J_2= 1/2$.  It is conspicuous in its; sudden spike and then
decay of $\chi_D$; sudden, discontinuous increase in $\chi_{120^{\circ}}$ and
$\chi_{\rho}$ and; drop and spike of $\chi_{CAF}$.  Such behavior is to be expected
due to the special 2-fold degenerate ground-state occurring {\it precisely} at the MG-point for a finite system.
For the $J_1-J_2$ chain this point is the one we previously identified using the fidelity $F_0$.
$\chi_{120^{\circ}}$ was constructed as a rough
probe of incommensurate or non-antiferromagnetic (i.e. $q \neq \pi$) ordering
and for $J_2>J_2^{MG}$ features develop in $\chi_{120^{\circ}}$ consistent with
incommensurate (short-range) correlations.
For $J_1-J_2$ chain it is known that \emph{short-range} incommensurate
correlations emerge at $J_2>J_2^{MG}$.  It
is noteworthy that the generalized fidelity susceptibility has sufficient
sensitivity to detect the onset of incommensurability effects beyond the Majumdar-Ghosh point. 
To summarize, for the $J_1-J_2$ it is clear that $\chi_{CAF}$ and $\chi_D$ detect
the quasi-AF and dimer order and at the same time the MG point is clearly identifiable with
the onset of incommensurability effects.

It is noteworthy that, as was discussed earlier, the MG point 
of the $J_1-J_2$ chain is connected, when tracked through
the $J'-J_2$ plane, with the unphysical parity transition of the anisotropic
nearest-neighbour triangular model.  
In particular since in the
isotropic triangular limit ($J'=1$, $J_2=0$) the system is known to exhibit
$120^{\circ}$ order and possess no excitation gap.  We therefore now turn our
attention to the $J_2=0$ anisotropic triangular lattice Heisenberg model.

\subsection{The ATHLM ($J_2=0$)}
\begin{figure}[t]
\includegraphics[scale=0.5]{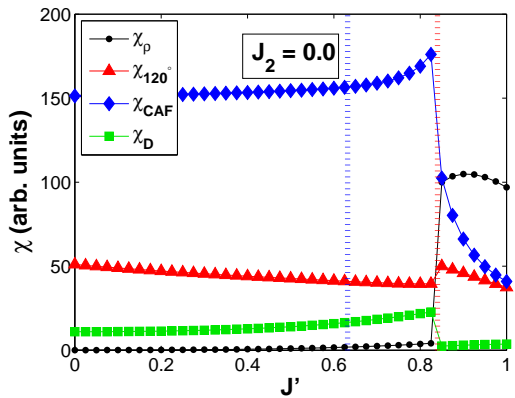}
\caption{\label{fig:FS_Plots_J2_0p0}  (Colour available on-line) Generalized
fidelity susceptibilities as a function of $J'$ for $J_2=0$ (i.e the ATLHM).
All marker, line, colour and scaling conventions are the same as those in
Fig.~\ref{fig:FS_Plots_Jp_0p0}. Results are for a $4\times 6$ system.}
\end{figure}

A plot of $\chi_{\rho}$, $\chi_D$, $\chi_{CAF}$ and $\chi_{120^{\circ}}$
($\delta \lambda = 10^{-4}$)  for $J_2=0$ for $J'<1$ can be found in
Fig.~\ref{fig:FS_Plots_J2_0p0}.  It is immediately apparent that there is again
a transition, corresponding to the downward spike in $F_0$ identified earlier,
at $J'=0.840=J'_c$ and that for $J'<J'_c$, $\chi_{CAF}$ and
$\chi_{120^{\circ}}$ behave in a qualitatively identical manner to the
Luttinger phase of the $J_1-J_2$ chain.  On the other hand, $\chi_D$ has no spike and
simply drops after the transition and although $\chi_{\rho}$ jumps abruptly to
a higher value at $J'_c$, it does not have a minimum anywhere in the $J'< 1$
region.  

In Ref.~\onlinecite{Thesberg_HATM_Twist} it was shown that the effect of
twisted boundary conditions, which allow for incommensurate correlations to
exist even in small finite systems, was to change the nature of this $J'_c$
transition from a parity transition to a first-order jump in the ground-state
ordering. This jump occurred at a lower $J'$ of 0.765 for $N=4 \times 6$ and it was observed that
incommensurate (short-range) spiral correlations persisted below this new transition though
the dominant interaction, and ground-state ordering, was consistent with
antiferromagnetism.  From the perspective of quantum fidelity susceptibilities used here
it is clear that collinear
antiferromagentic correlations are very important below the transition point, $J'<J'_c$. 
However, from the quantum fidelity susceptibilities alone we cannot rule at the existence
of a disordered state 
similar in character to that found in the $J_1-J_2$ chain for
$J_2<J_{2}^c$.  
We now turn to our results
for the generalized quantum fidelity susceptibilities in the
rest of the $J'-J_2$ plane (i.e. $J' \neq 0$, $J_2 \neq 0$) for the ANNTLHM.

\subsection{The ANNTLHM ($J_2=0$)}
\begin{figure*}[t]
\includegraphics[scale=0.55]{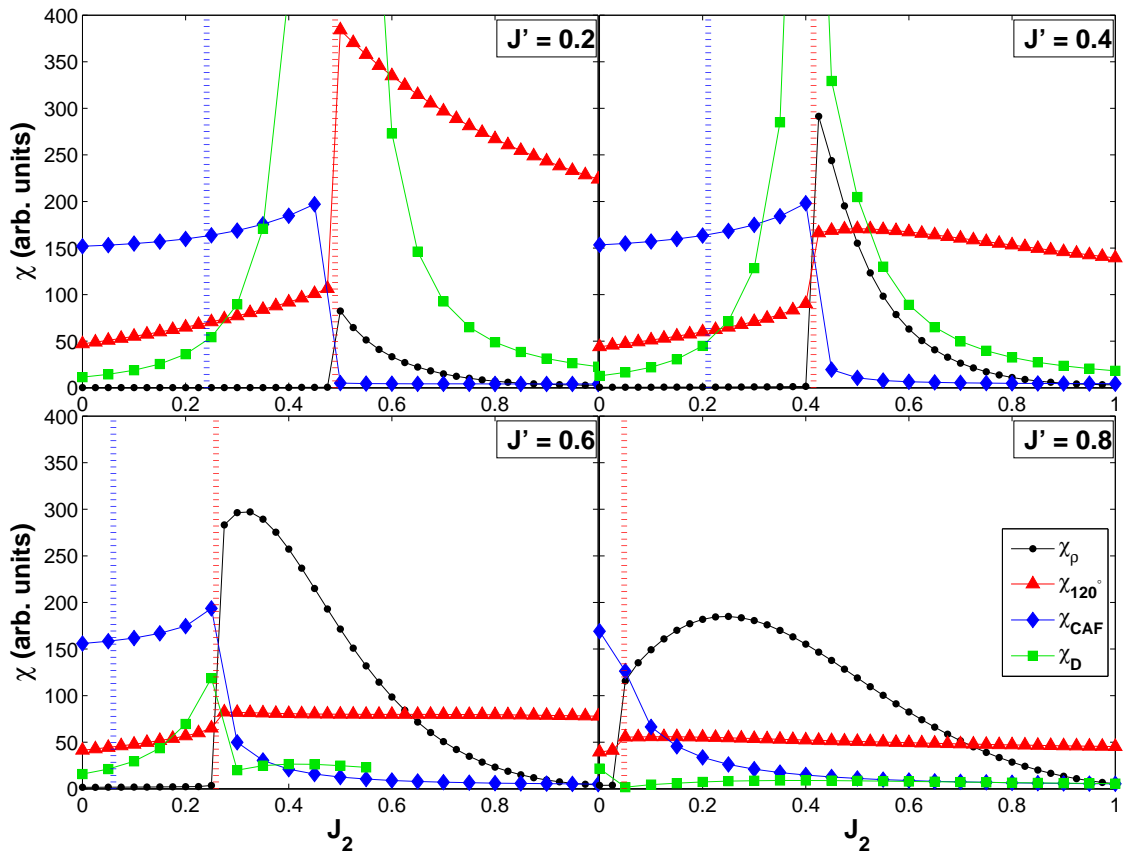}
\caption{\label{fig:FS_Plots_Other_Jp}  Generalized fidelity susceptibilities
as a function of $J_2$ for values of $J'=$ 0.2, 0.4, 0.6 and 0.8 (i.e.
cross-sections of the $J'-J_2$ plane).  All marker, line, colour and
scaling conventions are the same as those in
Fig.~\ref{fig:FS_Plots_Jp_0p0}. Results are for a $4\times 6$ system.}
\end{figure*}

\begin{figure}[h]
\includegraphics[scale=0.5]{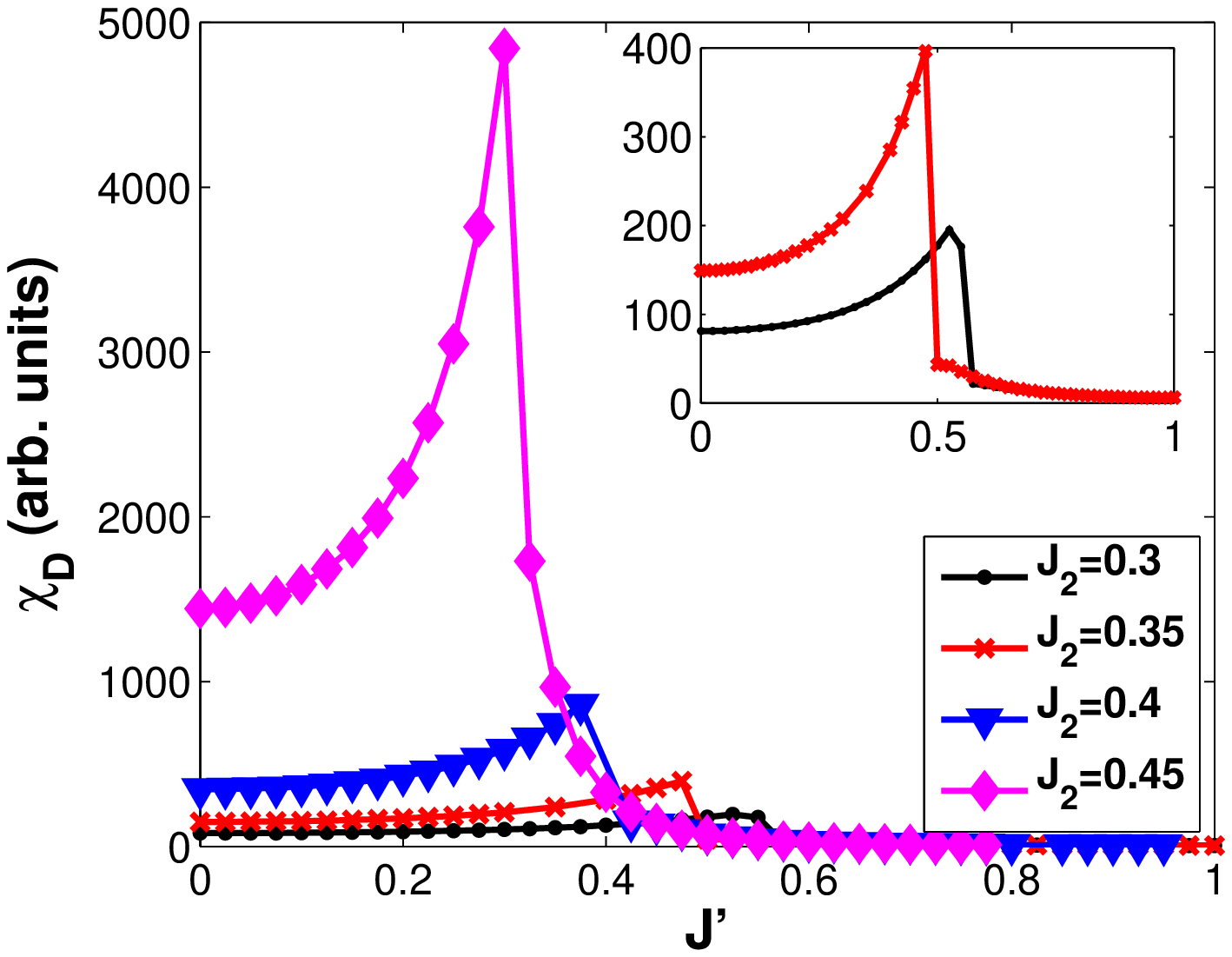}
\caption{\label{fig:Chi_D_Jp_Sweep}  $\chi_D$ as a function of $J'$ for $J_2=$
0.3 (black circles), 0.35 (red crosses), 0.4 (blue triangles) and 0.45 (magneta
diamonds).  The inset shows the same data for only $J_2=$ 0.35 and 0.3.  A
qualitative change in the nature of $\chi_D$
    can be seen at $J_2 \sim 0.4$.  For $J_2>0.4$ the peak in $\chi_D$ is significantly more pronouced.
    This could suggest a
transition from gapped dimer order with incommensurate short-ranged spiral
correlations to the true incommensurate spiral order that exists at
$J'=1,J_2=0$.}
\end{figure}

The same data gathered for the $J_1-J_2$ chain in
Fig.~\ref{fig:FS_Plots_Jp_0p0} is shown in Fig.~\ref{fig:FS_Plots_Other_Jp} for
the cases of $J'=$ 0.2, 0.4, 0.6 and 0.8.  These plots then serve to divide the
$J'-J_2$ plane into cross-sections in $J'$.  Again, the point identified by $F_0$ is clearly 
visible. Of note in these plots is the
consistent behaviour of $\chi_{\rho}$, $\chi_{120^{\circ}}$ and $\chi_{CAF}$ as
$J'$ increases lending evidence to the notion that the $J' < J'_c$ phase is
directly related to the $J_2<J_{2}^c$ phase.  The one marked difference is in
the behaviour of $\chi_D$ whose peaked nature becomes substantially less
pronounced as $J'$ grows.  This is indicative of a necessary (since they have
    different symmetries) transition from dimer to spiral order.
Unfortunately, there does not seem to be sudden features in $\chi_D$ vs. $J_2$
to identify this region.  A plot of $\chi_D$ vs. $J'$ for $J_2$s of 0.3 to 0.45
shown in Fig.~\ref{fig:Chi_D_Jp_Sweep}
does suggest a qualitative change in the way $\chi_D$ diverges at a
$J_2$ of approximately 0.4.  For $J_2>0.4$ the peak is much more pronouced than for $J_2<0.4$.
This could suggest a
transition from gapped dimer order with incommensurate short-ranged spiral
correlations to the true incommensurate spiral order. In Fig.~\ref{fig:Fidelity_Phase_Diagram}
this is indicated as the dotted red line.

As already stressed, the central observation to make from the results presented in Fig.~\ref{fig:FS_Plots_Other_Jp}
for $J',J_2\ll 1$ is the similarity with the results in Fig.~\ref{fig:FS_Plots_Jp_0p0} for $J_2<J_2^c$.
The presence of a non-zero $J'$ thus only changes the ordering in a very subtle way and possibly not at all.

\subsection{Non-collinear Versus Collinear Order ($\chi_{NCAF}$ vs. $\chi_{CAF}$)}

A final issue of interest is the competition between non-collinear and
collinear antiferromagnetic correlations in the anisotropic nearest-neighbour
triangular lattice.  Renormalization group
studies\cite{Balents_RG,Kallin_Ghamari_RG} of the triangular system suggest
that the $J' \ll 1$ phase is ordered antiferromagnetically and that crucial to
this ordering is the emergence of antiferromagnetic correlations between
\emph{next-nearest chains}.  In Ref.~\onlinecite{Thesberg_HATM_Twist} it was
found that, although next-nearest chain interactions were indeed of great
importance within that phase, there is intense competition between collinear
(CAF) and non-collinear (NCAF) ordering and that CAF is indeed the dominant
correlation, but only by an extremely small margin.  To re-investigate this claim a
separate generalized fidelity, $\chi_{NCAF}$, was constructed such that
next-nearest chains have ferromagnetic interactions and the two ($\chi_{CAF}$
    and $\chi_{NCAF}$) were computed for $J_2=0$, $J'<1$.  The field defining $\chi_{NCAF}$
is shown in Fig.~\ref{fig:Chi_Diagrams}.
As was the case in
Ref.~\onlinecite{Thesberg_HATM_Twist} the difference between the two is found
to be extremely small but $\chi_{CAF}$ is larger by a factor of approximately
0.001$\%$.   This minuscule discrepancy, though well within the realm of
numerical precision, suggests that the competition between these two types of
antiferromagnetic correlations is extremely fierce, at least within finite-size
systems.

\section{Conclusion}

In this paper the ground-state and excited-state quantum fidelities were used to track the
behaviour of the MG/Lifshitz point and BKT-type transition, found in the
$J_1-J_2$ ($J'=0$) chain, into the $J'-J_2$ plane.  It was found that both points trace
bounded regions within $J'-J_2$ plane and ultimately terminate on the $J'$ axis ($J_2=0$) corresponding to 
the anisotropic triangular Heisenberg model.  Specifically, the MG
point, which occurs as a ground-state level crossing in the $J_1-J_2$ chain
which is known to not survive in the thermodynamic limit, is connected to the
unphysical parity transition observed in the $J' < 1$ region of the anisotropic
triangular model. However, the region defined by the behavior of $F_1$ connecting the BKT transition
of the $J_1-J_2$ chain ($J'=0$) with a point on the $J'$ axis is strongly suggestive of
a new distinct phase.

In order to further explore and identify these phase regions, the generalized
fidelity susceptibilities $\chi_{\rho}$, $\chi_{120^{\circ}}$, $\chi_D$ and
$\chi_{CAF}$ were constructed. They are associated with the spin stiffness,
  $120^{\circ}$ spiral phase order parameter, dimer order parameter and
  collinear antiferromagnetic order parameter respectively.  These quantities
  are believed to be very sensitive and therefore well suited for finite system studies.  

When plotting these quantum fidelity susceptibilities  within the $J'-J_2$ plane 
the region defined by $F_0$ is readily identifiable while the $F_1$ region is much
more subtle. In the $J', J_2\ll 1$ region the $\chi_{CAF}$ is marginally favored over
$\chi_{NCAF}$ but it is not possible to conclusively eliminate the possibility of a disordered phase. 
Furthermore, the region above this phase (i.e. $J_2 > J_{2}^c$, $J' > J'_c$) is spiral ordered within a $N= 4 \times 6$ system.
This is known to be the case in the thermodynamic limit for the anisotropic
nearest-neighbour triangular model but for the $J_1-J_2$ chain this is known to
be false and for $J_2$ beyond the MG point incommensurate correlations are
only short-ranged for the $J_1-J_2$ chain.  For $J_2$ greater than approximately $0.4$ dimer correlations appear dominant. 
A possible way to distinguish these two phases would be
through a study of larger system sizes (like those done by Weichselbaum and
    White in Ref.~\onlinecite{Weichselbaum_and_White_DMRG}) to track the
closure of the energy-gap in the incommensurate phase of the $J_1-J_2$ chain as
that phase connects with the spiral-ordered phase of the triangular lattice
through the $J'-J_2$ plane.

An additional aspect not explored in this paper, due to the lack of available
system sizes, is the scaling behaviour of these generalized fidelity
susceptibilities throughout the $J'-J_2$ plane.  Such a study, potentially
viable through DMRG of a finite-cluster, would be very valuable and further solidify the
understanding of this phase diagram.

\begin{acknowledgments}

The authors would like to thank 
Sung-Sik Lee, Catherine Kallin and Sedigh Ghamari for many fruitful discussions.
We also acknowledge computing time at the Shared Hierarchical Academic
Research Computing Network (SHARCNET:www.sharcnet.ca) and research
support from NSERC.
\end{acknowledgments}

\bibliography{References}

\end{document}